\documentclass{PoS}

\def\microns{\mbox{ } \mu \mbox{m}}

\title{{\it Herschel} observations of {\it INTEGRAL} supergiant High Mass X-ray Binaries}

\ShortTitle{{\it Herschel} observations of {\it INTEGRAL} supergiant High Mass X-ray Binaries}

\author{\speaker{Sylvain Chaty}%
        \thanks{
We thank the organisers for a successfully organized and interesting workshop in the Bibliothèque Nationale de France.
This work is supported by the Centre National d'Etudes Spatiales (CNES), based on observations obtained with MINE -- the Multi-wavelength {\it INTEGRAL} NEtwork --.
}\\
       AIM (UMR-E 9005 CEA/DSM-CNRS-Université Paris Diderot)
       Irfu/Service d'Astrophysique, Centre de Saclay,  Bât. 709, FR-91191 Gif-sur-Yvette Cedex, France\\
       Institut Universitaire de France, 103, bd Saint-Michel 75005 Paris, France \\
       E-mail: \email{chaty@cea.fr}}

\author{Alexis Coleiro, Mathieu Servillat\\
       AIM (UMR-E 9005 CEA/DSM-CNRS-Université Paris Diderot)
       Irfu/Service d'Astrophysique, Centre de Saclay,  Bât. 709, FR-91191 Gif-sur-Yvette Cedex, France\\
        E-mail: \email{alexis.coleiro@cea.fr, mathieu.servillat@cea.fr}}

\author{Juan Antonio Zurita Heras \\
        Fran\c{c}ois Arago Centre, APC (Universit\'e Paris Diderot, CNRS/IN2P3, CEA/Irfu, Observatoire de Paris), 10 rue Alice Domon et L\'eonie Duquet, 75205 Paris Cedex 13, France \\
        E-mail: \email{zuritah@apc.univ-paris7.fr}}

\author{Farid Rahoui\\
        ESO, Karl-Schwarzschild-Strasse 2, D-85748 Garching bei München, Germany
\\
       E-mail: \email{frahoui@eso.org}}

\abstract{We present preliminary results on {\it Herschel}/PACS mid/far-infrared photometric observations of {\it INTEGRAL} supergiant High Mass X-ray Binaries (HMXBs), with the aim of detecting the presence and characterizing the nature of absorbing material (dust and/or cold gas), either enshrouding the whole binary systems, or surrounding the sources within their close environment.
These unique observations allow us to better characterize the nature of these HMXBs, to constrain the link with their environment (impact and feedback), and finally to get a better understanding of the formation and evolution of such rare and short-living supergiant HMXBs in our Galaxy.
}

\FullConference{ An INTEGRAL view of the high-energy sky (the first 10 years) - 9th INTEGRAL Workshop and celebration of the 10th anniversary of the launch\\
		 15-19 October 2012\\
		 Bibliotheque Nationale de France, Paris, France}

\begin{document}

\section{Introduction}

The high-energy {\it INTEGRAL} observatory, observing in the 20\,keV$-$8\,MeV range, has performed a detailed survey of the Galactic plane, and the ISGRI detector on the IBIS imager has discovered a new population of high energy binary systems, reported in \cite{bird:2010}. Most of these new sources are high mass X-ray binaries (HMXBs) hosting a neutron star orbiting a supergiant O/B companion. They are highly absorbed, exhibiting a huge intrinsic and local extinction, like for instance the extremely absorbed source IGR\,J16318$-$4848 \cite{filliatre:2004}. More details about the nature of these sources are reported in Chaty (\#013 in this volume of proceedings).

These systems, born with two very massive components, are likely the primary progenitors of neutron star/black hole binary mergers, probably related with short/hard gamma-ray bursts, and good candidates of gravitational wave emitters. It therefore appears that, because these obscured X-ray binaries represent a different evolutionary path of X-ray binaries, their study is of prime importance, providing hints on the formation and evolution of high energy binary systems, and a better understanding of the evolution of such rare and short-living high mass binary systems.

These sources are concentrated in a direction tangent to the Norma arm of the Galaxy \cite{chaty:2008}, a region of our Galaxy rich in star forming regions. As dust is a strong tracer of star formation, one of {\it Herschel}'s greatest strengths is the possibility to study the history of star formation in our Galaxy. This makes {\it Herschel} an enormously powerful facility to observe such obscured sources, giving a new insight for studying the process of formation and evolution of very massive stars in molecular clouds. These {\it Herschel} observations are therefore very valuable in the study of the latest phases of stellar evolution, particularly circumstellar shells, mass-loss in general, and stellar winds. They will also allow us to relate the properties of these objects to their birth environment.

\section{Observations and preliminary results}

We have performed sensitive far-infrared ($60-130 \microns$) imaging photometric observations with {\it Herschel}/PACS of 6 targets (see Table 1), some of them having the particularity to be the most absorbed supergiant High Mass X-ray Binaries of our Galaxy \cite{chaty:2008}. Observations were performed in the 3 available bands (blue 60-85\,$\microns$, green 85-130\,$\microns$ and red 130-210\,$\microns$) in mini-scan map mode (medium speed, 10 scan legs of 2.5' length with 2.0'' cross-scan step, with orientation angles at 70 and 110 degrees, and a repetition factor of 5 and 10 for each orientation angle for the blue/red and green/red filters respectively). We reprocessed the data with HIPE and custom scripts (M. Sauvage, priv. comm.), and we used the software {\it getsources} to extract fluxes \cite{menshchikov:2012}.

\begin{figure}
  \centerline{\includegraphics[width=15.cm]{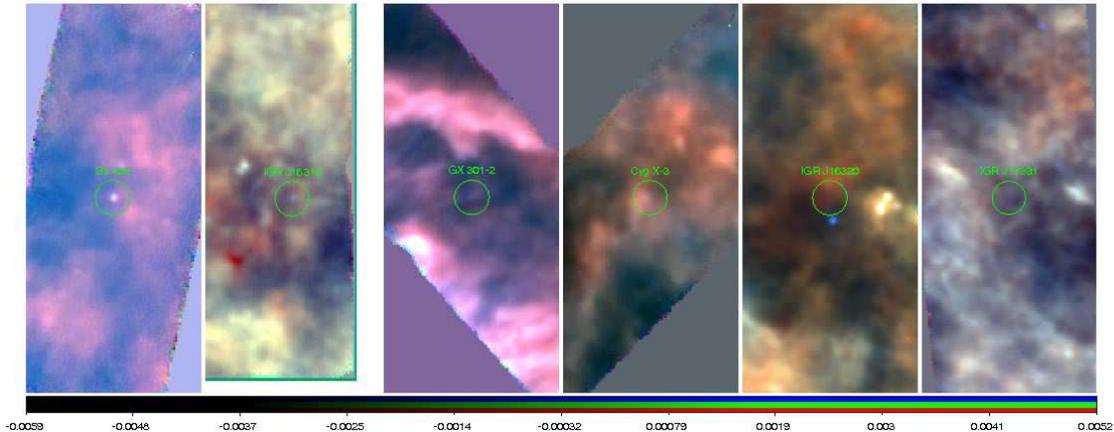}}
\caption{\label{figure:herschel}
{\it Herschel}/PACS observations of the 6 targets: SS\,433, IGR\,J16318-4848, GX\,301-2, Cyg\,X-3, IGR\,J16320-4751 and IGR\,J17391-3021 from left to right respectively. We point out that the blue point close to IGR\,16320-4751 is not associated to the X-ray source.
}
\end{figure}

\begin{figure}
  \centerline{\includegraphics[width=10.cm]{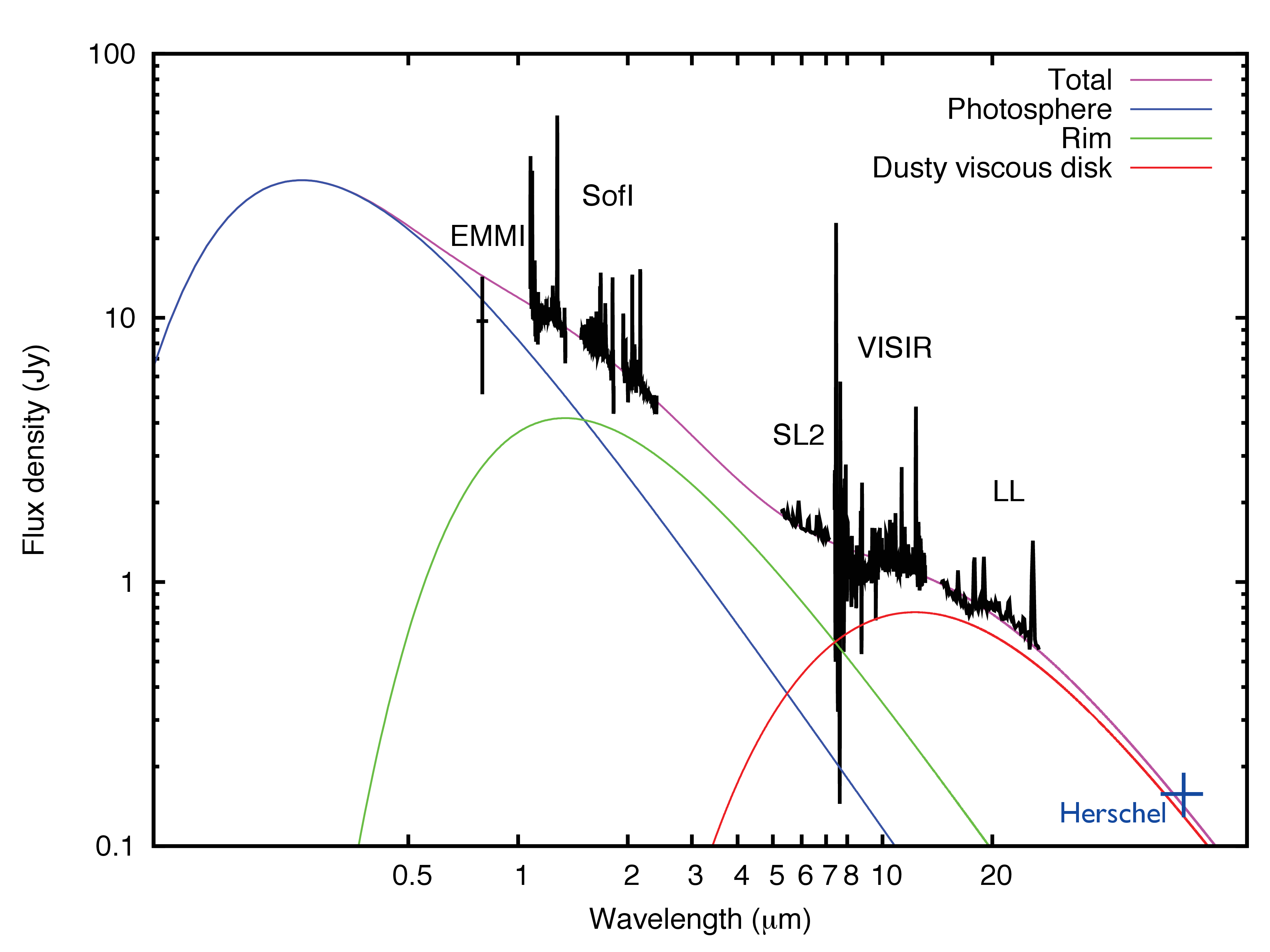}}
\caption{\label{figure:16318}
Broadband optical to mid-infrared (0.9 to 35 $\mu$m) ESO/NTT+VISIR and {\it Spitzer} reddened spectrum of IGR\,J16318$-$4848. Observations, fitted with both a model of a sgB[e] star and simple spherical dust component, require an extra (e.g. dust) component in order to fit its SED \cite{chaty:2012, rahoui:2008}. {\it Herschel} observations of this source at 70\,$\mu m$ (blue point) are consistent with the predictions of the star-forming model.
}
\end{figure}

The photometry results are reported in Table 1, and the preliminary images are shown in Figure \ref{figure:herschel}. We clearly detect the central sources in the fields of SS\,433, IGR\,J16318-4848 and GX\,301-2.
These far-infrared observations, the first ever performed in the far-infrared domain on high energy binary sources, allow us to study the point source emission, the impact of the source on its environment, and the feedback of the source on its environment. They complement multi-wavelength observations that we have already performed on these sources, with various ground-based and satellite facilities, allowing us to complete the broadband spectral energy distribution of these sources. For instance, we report in Figure \ref{figure:16318} the spectral energy distribution of the source IGR\,J16318-4848 derived in \cite{chaty:2012}, by using a model built for forming stars, i.e. a torus of dust and/or cold gas surrounding the supergiant star sgB[e]. We point out that the flux at 70\,$\mu m$ derived with {\it Herschel} (see Table 1) is consistent with the prediction of this model, confirming the model used in this paper, up to the far-infrared domain. Furthermore, these new data will allow us to adjust the model, and to make our predictions more accurate.

In addition, these far-infrared images suggest the presence of potential cavities detected in the surroundings of the central source, especially in the images of GX\,301-2 (see Figure \ref{figure:herschel}), suggesting that the strong radiation emanating from this source has created an empty space around the central source. Further analysis, by e.g. HI and extinction measurements, has to be performed in order to check whether the surrounding gas is located at the same distance than the central source.
A complete analysis of these {\it Herschel} observations for each of these high energy sources is ongoing, and papers are currently being prepared.

Table 1. List of targets, with RA DEC position, spectral type (Sp. T.), distance (Dist.), and flux at 70\,$\mu m$, 100\,$\mu m$ and 160\,$\mu m$ respectively. \\
\scriptsize  
\begin{tabular}{c c c c c c c c}
\hline\hline       
Source & R.A. (J2000.0) & Dec. (J2000.0) & Sp.T. & Dist. [kpc] & F$_{70 \, \mu m}$ (mJy) & F$_{100 \, \mu m}$ (mJy) & F$_{160 \, \mu m}$ (mJy) \\
\hline                    
GX 301$-$2 & 12:26:37.599 & -62:46:14.005 & B1--1.5 Ia & 4.1 & 61 $\pm$ 7 & 61 $\pm$ 15 & - \\
IGR J16318$-$4848 & 16:31:48.6 & -48:48:59.998 & sgB[e] & 1.6 & 140 $\pm$ 14 & 122 $\pm$ 39 & - \\
IGR J16320$-$4751 & 16:32:01.901 & -47:52:27.005 & BN0.5Ia & 3.5 & - & - & - \\
IGR J17391$-$3021 & 17:39:11.58 & -30:20:37.604 & O8 Iab(f) & 2.7 & - & - & - \\
SS 433 & 19:11:49.57 & +04:58:57.9 & A7Ib & 5.5 & 101 $\pm$ 2 & 100 $\pm$ 3 & 94 $\pm$ 35 \\
Cyg X-3 & 20:32:25.78 & +40:57:27.9 & WNe$+$ & 9.0 & - & - & - \\
\hline
\end{tabular}
\normalsize

\section{Conclusions}

We report here that: 
{\it i.} {\it Herschel} has detected the far-infrared counterpart of three X-ray sources;
{\it ii.} These new data confirm the presence of a disk of dust around IGR\,J16318-4848, and will lead to a similar analyis for the sources GX\,301-2 and SS\,433;
and {\it iii.} the new images suggest the presence of cavities around some sources, such as GX\,301-2, leading to the study of the impact of X-ray sources on their environment, but also of the feedback of these sources on the ISM.

These new {\it Herschel} observations will allow us to study these absorbed supergiant X-ray binaries, the surrounding gas component, and where it comes from, by better characterizing its properties, such as the temperature, composition, geometry, extension around the system, etc... In addition, adding {\it Herschel} detections on broad-band spectral energy distributions should reveal the presence of an extended circumstellar dust and/or cold gas component around some of these sources. The origin of this component is either related to the birth of the progenitor of such supergiant stars, their nucleosynthesis, or due to stellar wind photoionization. These observations will therefore allow us to answer the most important questions: Is such an unusual circumstellar environment due to the stellar evolution OR to the binarity of the system itself? Does the supergiant star influence its close environment, for instance by triggering additional stellar formation, and/or does the interstellar medium influence the star, by a feedback effect? These questions are fundamental issues, because these sources are also the progenitors of the extreme and poorly known Luminous Blue Variable (LBV) stars, and dusty wind hypergiant stars.

\end{document}